\documentclass[aps,prd,twocolumn,superscriptaddress,showpacs,amsmath,amssymb]{revtex4}

\usepackage{xspace}
\usepackage{dcolumn}
\usepackage{bm}
\usepackage{setspace}
\usepackage{graphicx}
\usepackage{amssymb}
\usepackage{epsfig}

\newenvironment{cfigure1c}[1][tbp]{\begin{figure*}[#1]\centering}{\end{figure*}}

\newcommand{\ttbar}  {\ensuremath{t\bar{t}}\xspace}

\newcommand{\et}     {\ensuremath{E_{T}}\xspace}

\newcommand{\pt}     {\ensuremath{p_{T}}\xspace}

\newcommand{\mtop}    {\ensuremath{{M}_{\mathrm{top}}}\xspace}

\newcommand{\mnwa}    {\ensuremath{m_{t}^{\mathrm{NW}}}\xspace}
\newcommand{\lpt}    {\ensuremath{p_{T}^{\mathrm{lepton}}}\xspace}

\newcommand{\mTT}    {\ensuremath{m_{\mathrm{T2}}}\xspace}

\newcommand{\djes}   {\ensuremath{\Delta_{\mathrm{JES}}}\xspace}

\newcommand{\gevcc}[1]  {\ensuremath{#1~\mathrm{GeV}/c^{2}}}

\newcommand{\invfb}[1]  {\ensuremath{#1~\mathrm{fb}^{-1}}}

\makeatletter
\AtBeginDocument{\@ifpackageloaded{natbib}{\ifNAT@numbers\if@filesw\immediate\write\@auxout{\string\global\string\NAT@numberstrue}\fi\fi}{}}
\makeatother
\begin{document}

        \title{Top-Quark Mass Measurement in the Dilepton Channel Using {\it in situ} Jet Energy Scale Calibration}

        \author{Hyun Su Lee}
        \affiliation{Department of Physics, Ewha Womans University, Seoul 120-750, Korea}
        \date{\today}

\begin{abstract}
        We employ a top-quark mass measurement technique in the dilepton channel with {\it in situ} jet energy scale calibration.  Three variables having different jet energy scale dependences are used simultaneously to extract not only the top-quark mass but also the energy scale of the jet from a single likelihood fit. Monte Carlo studies with events corresponding to an integrated luminosity of \invfb{5} proton-proton collisions at the Large Hadron Collider $\sqrt{s} = 7$~TeV are performed. Our analysis suggests that the overall jet energy scale uncertainty can be significantly reduced and the top-quark mass can be determined with a precision of less than \gevcc{1}, including jet energy scale uncertainty, at the Large Hadron Collider.

\end{abstract}

\maketitle
In the standard model, the top quark ($t$) is the heaviest known elementary particle~\cite{pdg}. The heavy mass significantly affects the electroweak radiative correction that relates the top-quark mass and the $W$ boson mass to the Higgs boson mass~\cite{elfit}. In addition, the heavy mass may have implications for new physics theories, including the minimal supersymmetric standard model and technicolor-like models. Because of its importance,  top-quark mass~(\mtop) measurements have been performed using various methods in different decay channels. The precision of the \mtop measurement already surpasses 0.5\% at the Tevatron~\cite{top_average}.

It is important to measure \mtop using different techniques and independent data samples in different decay channels. Significant differences in the measurements of \mtop using different decay channels may indicate contributions from new physics beyond the standard model~\cite{topnp}. The dilepton decay channel is particularly interesting because the signature of this channel can be mimicked with supersymmetric partner stop pairs~\cite{stop} as well as charged Higgs boson signals~\cite{chiggs}. However, the \mtop precisions of the dilepton channel measurements~\cite{massdilcdf,massdild0,massdilcms} have been limited because the branching fraction is much smaller and systematic uncertainty from the jet energy scale~(JES) is much larger than those of the lepton+jets channel measurements~\cite{massljcdf,massljd0,massljcms,massljatlas}. At the Tevatron, a single measurement already achieved \gevcc{1.2} precision in the lepton+jets channel~\cite{massljcdf}. However, the total uncertainty in the dilepton channel in a standalone measurement was \gevcc{3.0}~\cite{massdild0}, with the uncertainty from JES being dominant.
In the Large Hadron Collider~(LHC), the cross section of \ttbar pair production is approximately 20 times larger than that in the Tevatron. Therefore, the small branching fraction corresponding to data obtained in LHC experiments up to 2011 (more than \invfb{5} integrated luminosity) is not an issue. As an example, the recent \mtop measurement in the dilepton channel performed by the CMS Collaboration had \gevcc{1.2} statistical uncertainty using \invfb{2.2} data~\cite{massdilcms}. However, this measurement returned a larger systematic uncertainty,~(\gevcc{^{+2.5}_{-2.6}}), which is dominated by the overall JES uncertainty~(\gevcc{2.0}). The JES systematic uncertainty cannot generally be reduced by using a larger data sample.

A similar issue was encountered in the lepton+jets channel. However, {\it in situ} JES calibration using hadronic decaying $W$ bosons~\cite{tmt1} resolved this issue. In this method, the overall JES uncertainty was absorbed in the statistical uncertainty; therefore, a larger data sample could reduce not only the pure statistical uncertainty but also the overall JES systematic uncertainty.
However, in the dilepton channel, no single variable can be used to calibrate the JES uncertainty {\it in situ} for the \mtop measurement. A recent measurement performed by the D0 Collaboration employed a new technique for JES calibration in this channel~\cite{d0dilnew}. They used the result of the JES measurement in the lepton+jets channel. In this way, they significantly reduced overall JES systematic uncertainty in the dilepton channel. However, it is not a standalone measurement and the use of the lepton+jets channel result introduced an additional systematic uncertainty.

In this Letter, we propose a novel technique for the \mtop measurement in the dilepton channel, in which we perform the {\it in situ} JES calibration using three variables. The selected variables, which have already been used for the \mtop measurements, have different dependences on \mtop and JES, so we can perform simultaneous measurement of both \mtop and JES.
This is the first time  the \mtop measurement has been performed using {\it in situ} JES calibration in the dilepton channel even though it is a Monte Carlo~(MC) simulated experiment.
For a realistic estimation of the precision, we consider the environment of the LHC experiment with \invfb{5} $pp$ collisions.

There are different definitions of quark masses in the theoretical framework~\cite{mass_theory}. \mtop of the $\overline{\mathrm{MS}}$ renormalization scheme, $M_{\overline{\mathrm{MS}}}$, differs from the pole mass, $M_{\mathrm{ploe}}$, by about \gevcc{10}~\cite{mass_theory,pole_mass}. The direct \mtop measurements are all calibrated with MC simulations; the measured quantities therefore  correspond to the \mtop scheme used in the MC simulations, $M_{\mathrm{MC}}$. Even though it is usually assumed that $M_{\mathrm{MC}}$ is the same as $M_{\mathrm{pole}}$~\cite{top_average}, there are a number of theoretical questions in relating $M_{\mathrm{pole}}$ to $M_{\mathrm{MC}}$~\cite{mass_theory,mass_theory1}. An accurate relation is still under theoretical investigation~\cite{mass_theory2}.
As in all other direct \mtop measurements,  we measure $M_{\mathrm{MC}}$ in this Letter.

We generate simulated \ttbar samples using the leading order~(LO) MC generator {\sc madgraph/madevent} package~\cite{madgraph}. In the {\sc madgraph/madevent} generation, we vary the parameter  \mtop between 160  and \gevcc{190} in steps of \gevcc{2}. A total of 500~000 \ttbar pair events in the dilepton final state are produced for each sample. Showering and hadronization are performed by {\sc pythia}~\cite{pythia}. To take into account the detector effect and perform event reconstruction under realistic conditions, we use the fast detector simulation package~{\sc delphes}~\cite{delphes}. The conditions used for the simulated detector are the ones generally associated with the LHC detector.  We assume the coverage of the tracker to be within $|\eta|=2.5$ with 100\% efficiency and that of the calorimeter to be within $|\eta|=3$ with tower segment $\Delta\eta \sim 0.1$ and $\Delta\phi \sim 0.1$. The resolution of the electromagnetic calorimeter~(EM) and hadronic calorimeter~(HA) are parameterized by
$$\frac{\sigma_{EM}}{E}=0.005+\frac{0.25}{E}+\frac{0.05}{\sqrt{E}},$$
$$\frac{\sigma_{HA}}{E}=0.05+\frac{1.5}{\sqrt{E}}.$$
All physics objects such as leptons, jets, and missing transverse energy are reconstructed in the fast simulation. Jets originating from $b$ quarks are tagged using a reliable $b$-tagging algorithm with an efficiency of approximately 40\%.

In the dilepton decay channel, the production of a \ttbar pair is followed by the decay of each top quark to a $W$ boson and a $b$ quark, where both $W$ bosons decay to charged leptons~(electron or muon) and neutrinos~($\ttbar \rightarrow ll'\nu\nu'b\bar{b}$). Events in this channel thus contain two leptons, two $b$ quark jets, and two undetected neutrinos. To select the candidate events of \ttbar dilepton topology, we require two oppositely charged lepton candidates with $\pt > 20$~GeV/c. We also require a missing transverse energy exceeding 25~GeV and at least two tagged jets with $\et > 30$~GeV. The expected signal and background events are $5961 \pm 545$ and $320 \pm 97$, respectively, taken from Ref.~\cite{massdilcms} with scaling according to the respective integrated luminosity for the \invfb{5} data. The expected background contribution is approximately 5\%. Because of its small contribution, we do not include background events in our further analysis, for simplicity.

We use three reconstructed variables as observables of \mtop as well as the nuisance parameter \djes. \djes is the relative energy scale of the jet with respect to the nominal JES as used in the lepton+jets channel measurements~\cite{massljcdf,massljd0,massljcms,massljatlas}. All three variables have already been used in the \mtop measurements.
\begin{enumerate}
        \item \mnwa~-- The reconstructed top-quark mass obtained using the neutrino weighting algorithm~\cite{nwa}:  \mnwa is still widely used in the dilepton channel~\cite{massdilcdf,massdild0,massdilcms} as a template method. A study by the CDF Collaboration indicates that \mnwa has the most-precise statistical uncertainty among the variables used in its study~\cite{mT2CDF}.
        \item \mTT~-- The collider variable related to the missing transverse mass of the system of two missing particles: \mTT was initially developed for new physics particles as a variable sensitive to the mass of new particles in pair production~\cite{mT2}. Because of the two missing particles~(neutrinos) in the dilepton channel,  this variable was suggested for the \mtop measurement~\cite{mT2LHC}. The CDF Collaboration performed the \mtop measurement using \mTT and improved the precision with the simultaneous usage of \mnwa and \mTT~\cite{mT2CDF}. They also extensively studied the statistical as well as systematical uncertainties in the different variables. The expected statistical uncertainty obtained using \mnwa had the smallest value; however, the systematic uncertainty was larger than that in the measurement performed using \mTT. This difference was caused by different JES systematics due to the different dependence of each variable on \mtop and JES.
	\item \lpt~-- The average \pt of two leptons: \lpt has been suggested as a good variable for the \mtop measurement with large statistics in the LHC~\cite{lptLHC}. \lpt, in general, does not depend on JES, and it negligibly contributed to the systematic uncertainty from JES. The CDF Collaboration has performed measurements using this variable in the lepton+jets~\cite{lptCDF} and the dilepton~\cite{lptCDFdil} channels,  obtaining negligible JES systematic uncertainty but relatively large statistical uncertainty. This is basically caused by the low sensitivity of \mtop and the insensitivity of JES to the lepton \pt variable.
\end{enumerate}

\begin{cfigure1c}
\begin{tabular}{ccc}
\includegraphics[width=0.31\textwidth]{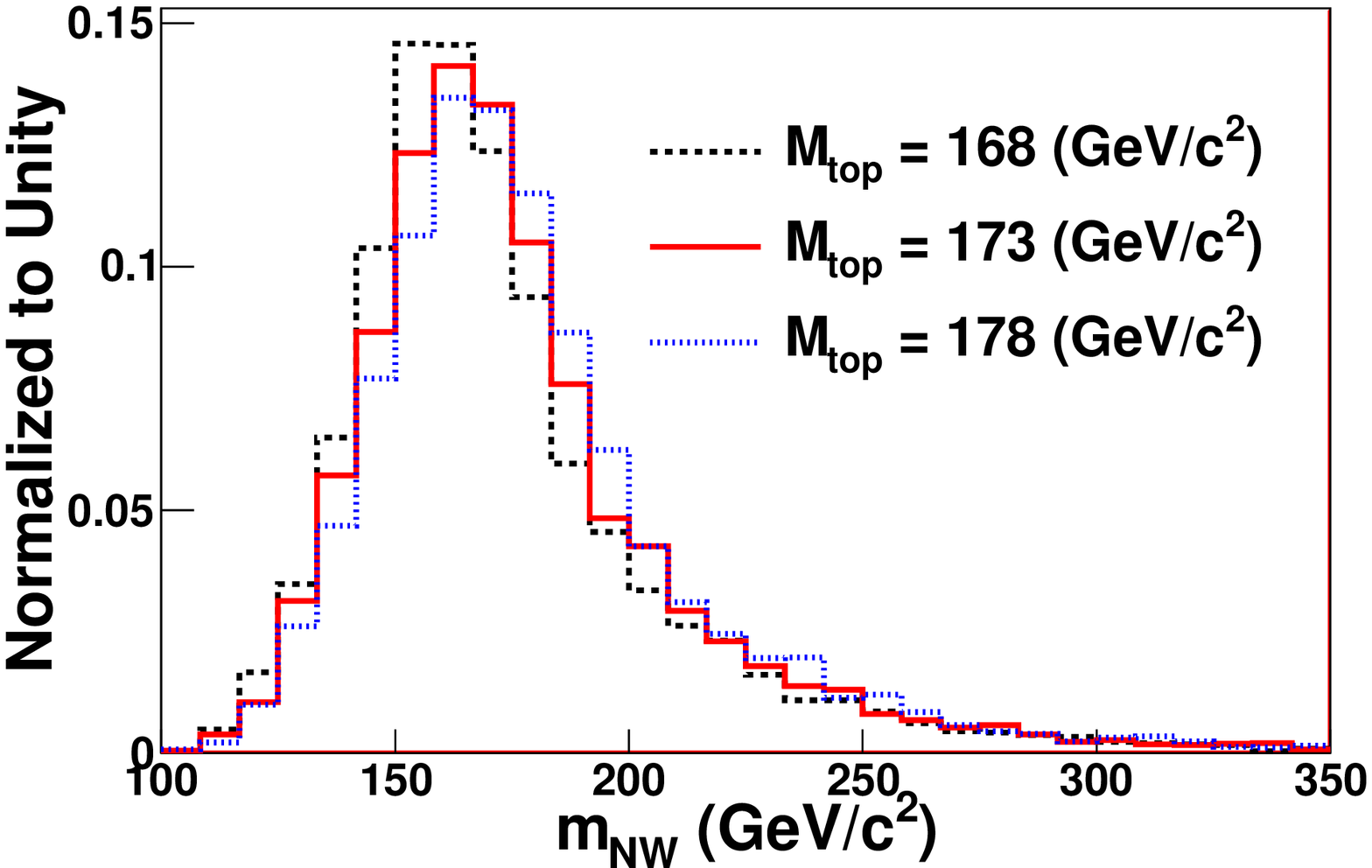} &
\includegraphics[width=0.31\textwidth]{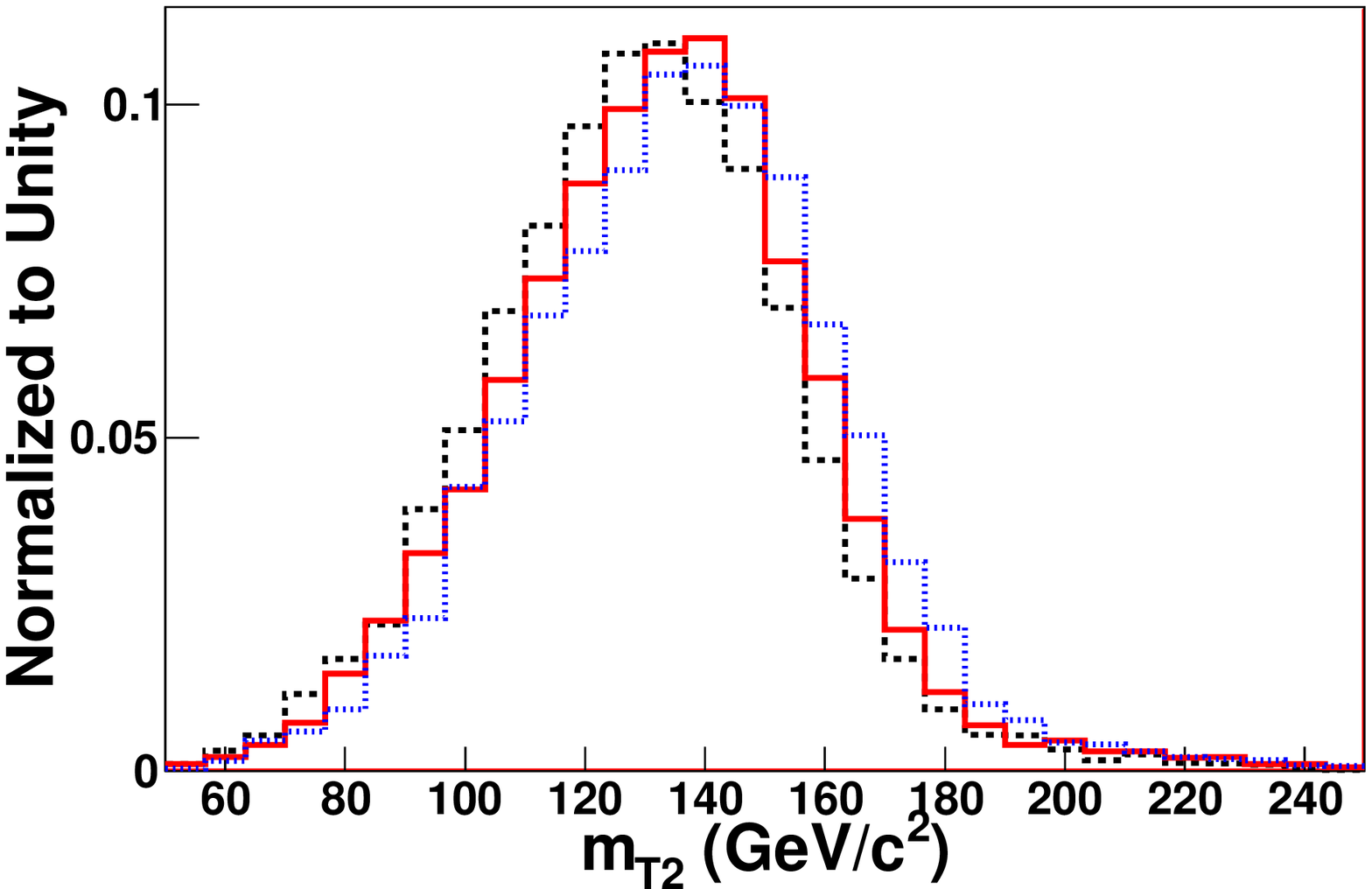} &
\includegraphics[width=0.31\textwidth]{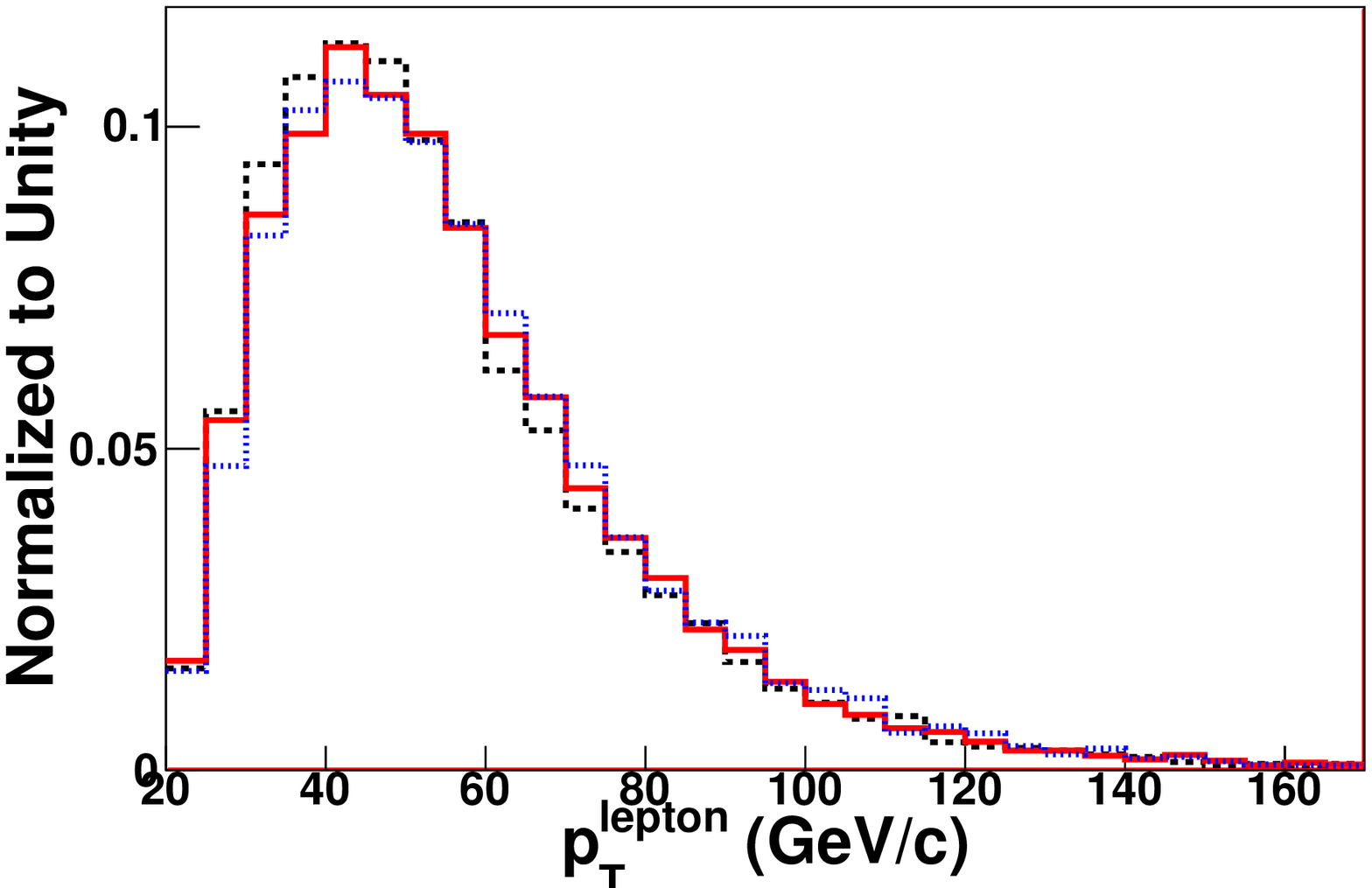} \\
(a) \mnwa &(b) \mTT &(c) \lpt  \\
\includegraphics[width=0.31\textwidth]{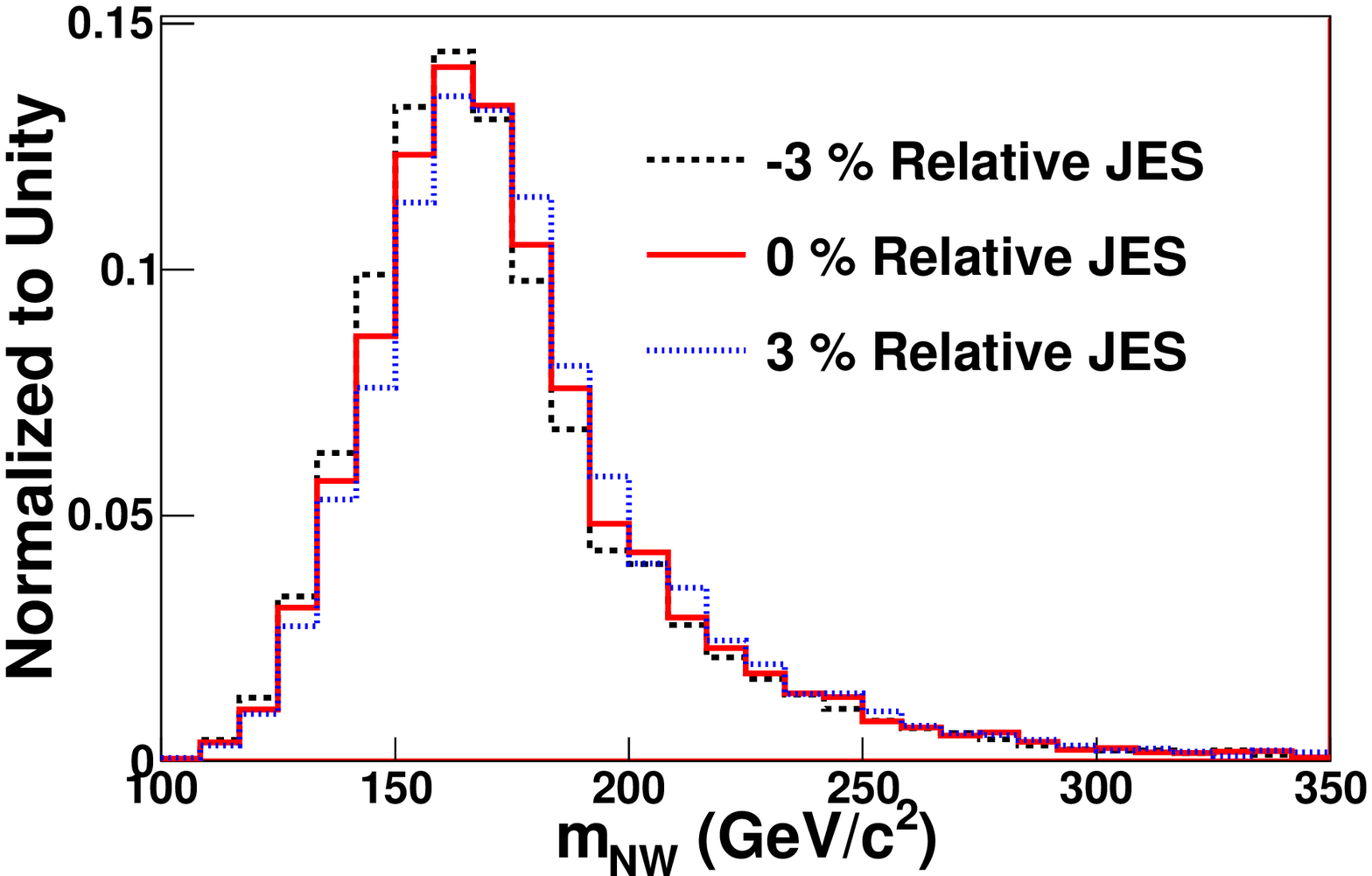} &
\includegraphics[width=0.31\textwidth]{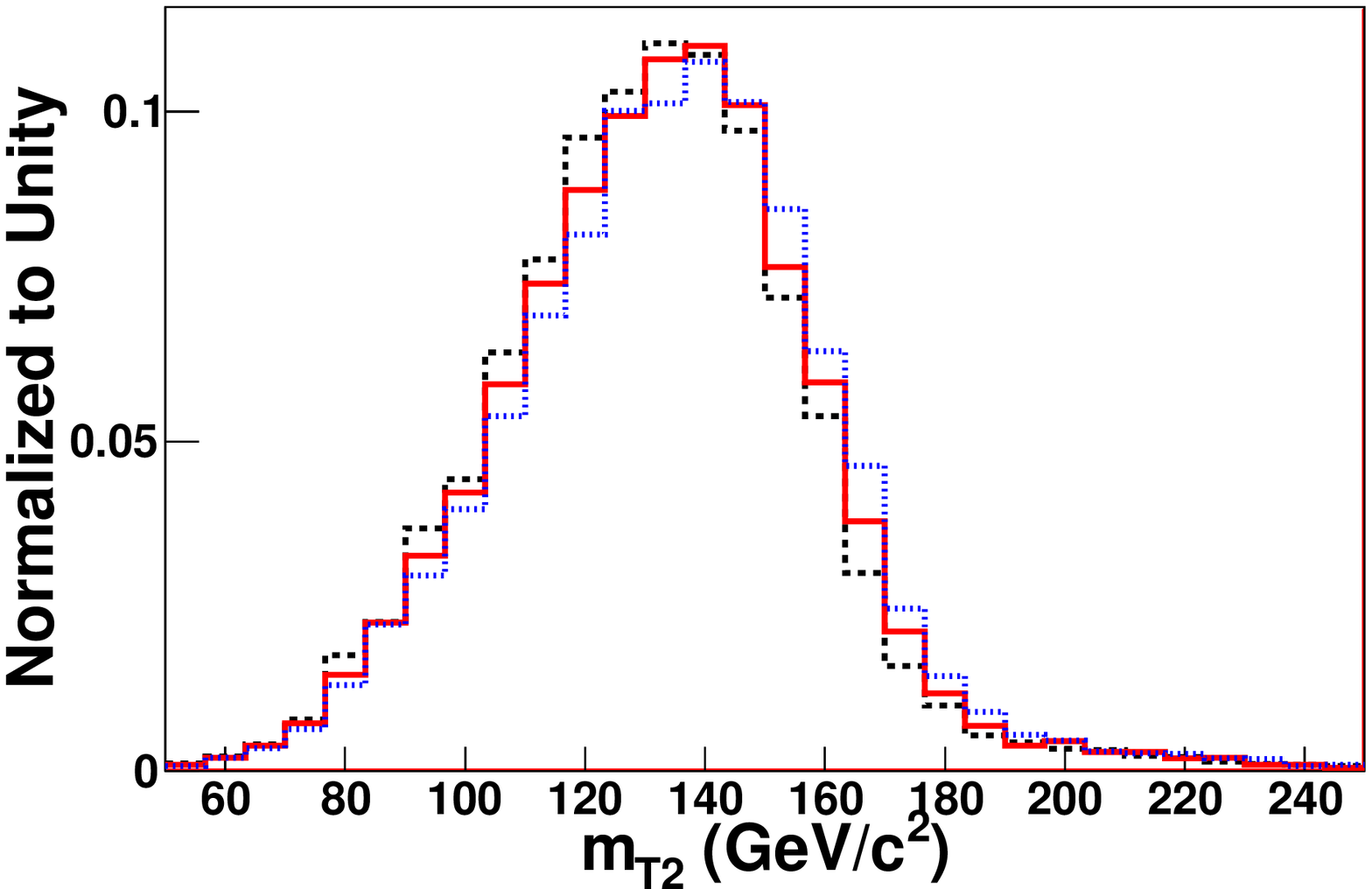} &
\includegraphics[width=0.31\textwidth]{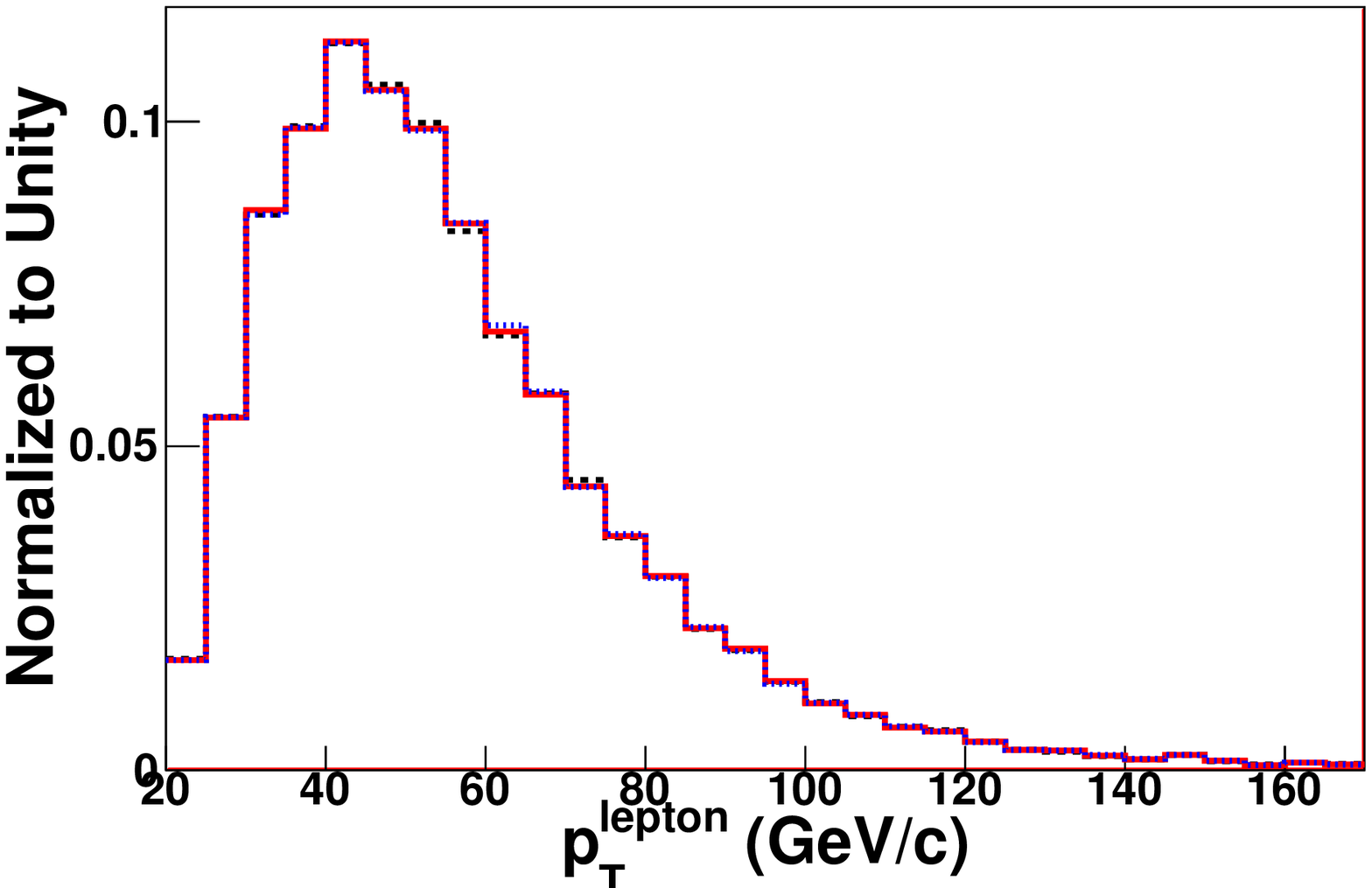} \\
(d) \mnwa &(e) \mTT &(f) \lpt \\
\end{tabular}
\caption[variable dependence]{Plots of the three variables for different \mtop~(a, b, c) and \djes~(d, e, f).}
\label{ref:compmass}
\end{cfigure1c}

The three variables discussed above have very different \mtop and JES dependences. Therefore, it is possible to extract both \mtop and JES information together if we use the three variables simultaneously. To simulate and extract JES information from the MC pseudoexperiments, we vary the scale of the jet energy relative to nominal JES~(\djes) in the MC simulations from $-$15\%  to 15\%  in 1.5\% steps. Even though the overall JES uncertainties depend on the jet $\eta$ and $\pt$, we use the overall variation for simplicity in this study. The overall jet energy uncertainty in the CMS experiment~\cite{cmsjes} is expected to be from 2\% to 3\%.
The plots of each variable are shown in Fig.~\ref{ref:compmass} with different \mtop as well as  \djes. \mnwa and \mTT depend on both \mtop and \djes. However, it is clear that \lpt does not depend on \djes.
Therefore, using three variables simultaneously in a single likelihood function, we can extract \djes information {\it in situ} as a nuisance parameter of the \mtop measurement.

The technique of using three variables in a single likelihood function was developed by the CDF Collaboration and has been used in lepton+jets channel measurements. We follow the procedure discussed in Ref.~\cite{massdilcdf}.
We estimate the probability density functions~(PDFs) of signals, $P_{sig}(\mnwa,\mTT,\lpt;\mtop,\djes$), using kernel density estimation~\cite{KDEHEP}. For the discrete values of \mtop and \djes, we estimate the PDFs for the observables. We smooth and interpolate the MC distributions to find PDFs for arbitrary values of \mtop and \djes using the local polynomial smoothing method~\cite{lps}. We then build the unbinned maximum
likelihood~\cite{eml} for $N$ events (and $n_{s}$ expected signal events) with the Poisson fluctuation:
$$\mathcal{L} = \frac{e^{-n_{s}}{n_{s}}^{N}}{N!}\Sigma_{i=1}^{N}P_{sig}(\mnwa,\mTT,\lpt;\mtop,\djes).$$
The quantity $P_{sig}(\mnwa,\mTT,\lpt;\mtop,\djes)$ denotes the signal PDFs as determined by kernel density estimation and local polynomial smoothing as a function of \mtop and \djes.
We minimize the negative logarithm of the likelihood using {\sc minuit}~\cite{minuit} with respect to all three parameters~($n_{s}$, \mtop, and \djes). The uncertainty on \mtop or \djes is found by searching for the points where the negative logarithm of the likelihood minimized with respect to all other parameters deviates by 1/2 from the minimum. The uncertainty on the \mtop measurement obtained in this way includes the statistical uncertainty as well as the systematic uncertainty of the overall JES owing to the allowed variation of \djes.  We also perform the likelihood fit without varying \djes~(\mtop-only fit) and set it to zero. The \mtop-only fit measurement allows us to compare the performance of the new technique with the {\it in situ} JES calibration~(2D fit).

\begin{figure}
\includegraphics[width=0.45\textwidth]{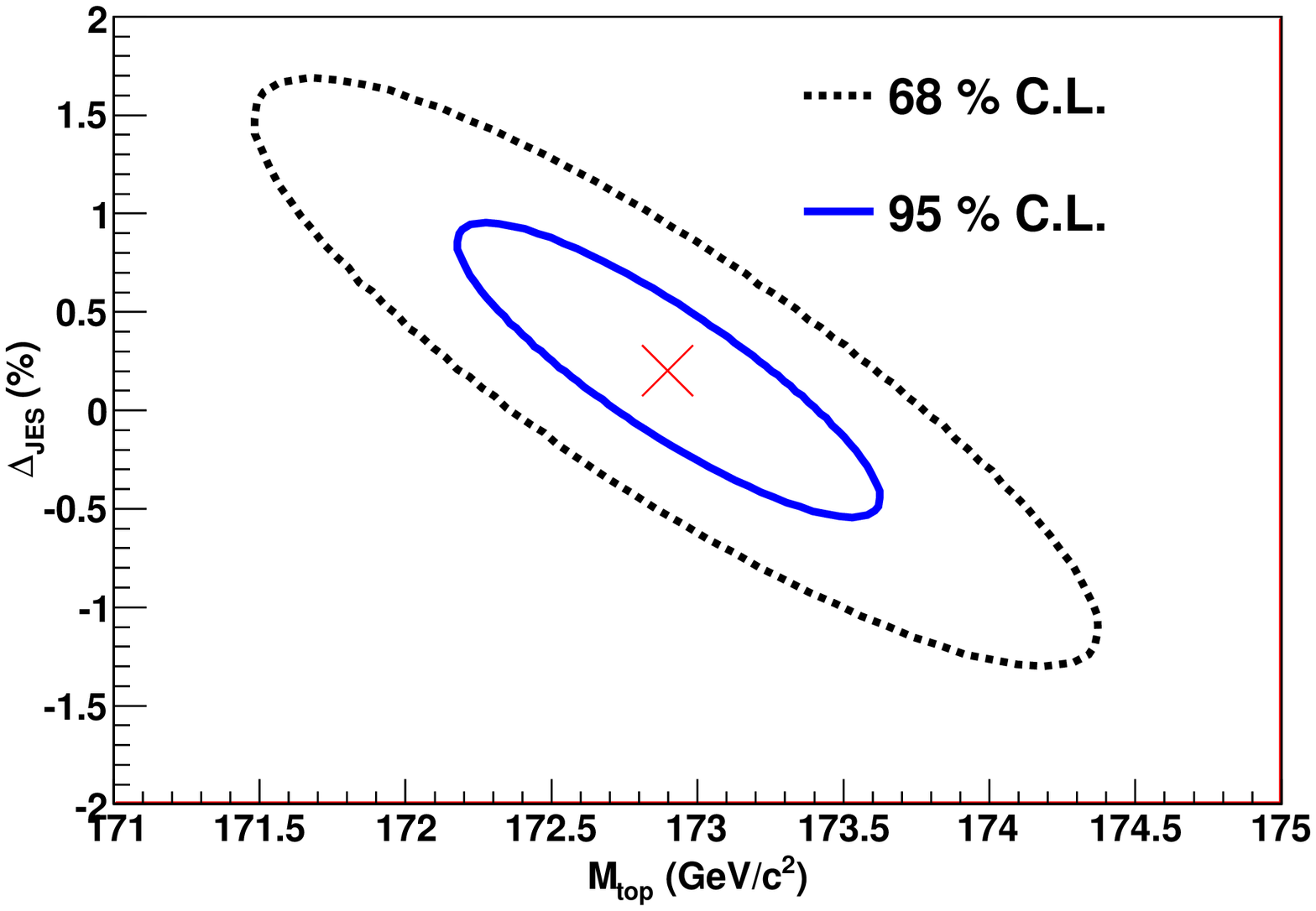}
\caption[variables dependence]{Negative log-likelihood contours of a single pseudoexperiment using a \mtop~=~\gevcc{173} and \djes~=~0.0\% sample. The minimum is indicated by the ``x'' and corresponds to the most probable top-quark mass and \djes. The contours are drawn at 68\% confidence level and 95\% confidence level, corresponding to 1$\sigma$ and 2$\sigma$ uncertainty of the measurement, respectively.}
\label{ref:contour}
\end{figure}

We test the likelihood procedure using MC pseudoexperiments. We construct pseudodata from a certain value of \mtop and \djes. We select the number of signal events from a Poisson distribution with a mean equal to the expected number of signal events, $5961 \pm 545$. We perform the maximum likelihood fit described in the previous section. In principle, the likelihood fit, on average, returns the value of the top quark mass used to generate the pseudoexperiments. Figure~\ref{ref:contour} shows an example of a likelihood fit contour from a single pseudoexperiment of the 2D fit. We use a \mtop~=~\gevcc{173} and \djes~=~0\% sample for this experiment and obtain \mtop~=~\gevcc{172.89\pm0.57} and \djes$=0.22 \pm 0.64\%$. We verify the \mtop-only fit using the same pseudodata and obtain \mtop~=~\gevcc{173.12\pm0.37}. We perform this experiment 3~000 times for seven different \mtop values ranging from {168} to \gevcc{178}. Figures~\ref{ref:bias_nocorr} (a) and (b) show the average residual~(deviation from input \mtop) and the width of the pull~(the ratio of the residual to the uncertainty reported by {\sc minuit}), respectively, in the 2D fit using samples of \djes$=0.0\%$ without corrections.
The small positive bias, \gevcc{0.43}, is corrected and the uncertainty is correspondingly increased by 5\% to correct for the width of the pull distribution. The residual and pull width for the \djes parameter are also investigated using the same  pseudoexperiments. We obtain a small negative bias of 0.33\%
and a pull width of 1.08. We apply suitable corrections for the nuisance parameter \djes. These corrections enable us to test the \mtop measurement with different \djes parameters. We vary the input \djes from $-$3.0\% to 3.0\%, corresponding to approximately 1 to 1.5 times the JES uncertainty in the LHC experiment~\cite{cmsjes}. Figure~\ref{ref:bias_nocorr} (c) shows the residual distribution for various \djes samples. There is no significant effect of  \djes on the mass residual. Therefore, we do not apply a correction of \mtop for the \djes parameter.
The same procedures for the \mtop-only fit are performed, and we find no bias. The width of the pull distribution is also consistent with unity.

\begin{cfigure1c}
\begin{tabular}{ccc}
\includegraphics[width=0.31\textwidth]{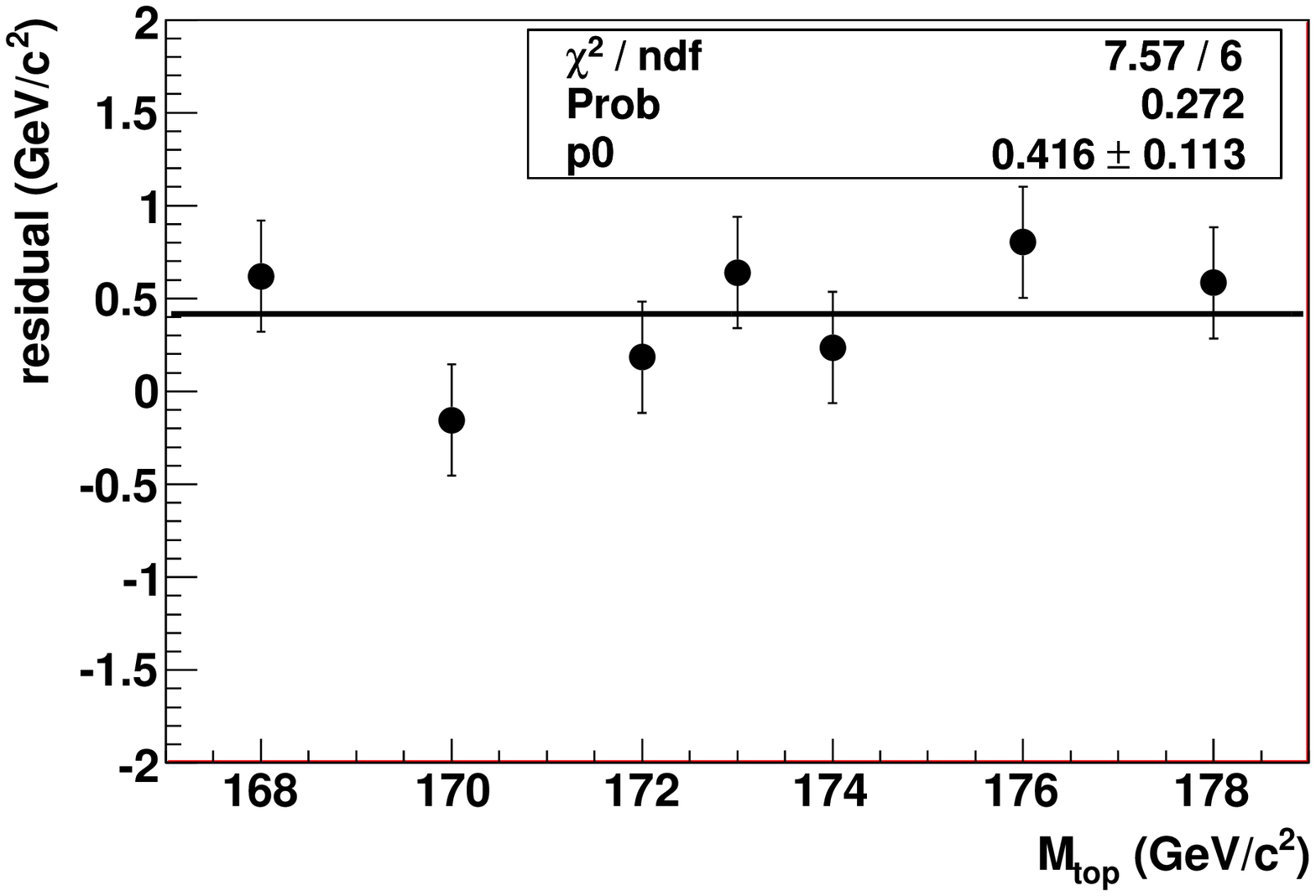} &
\includegraphics[width=0.31\textwidth]{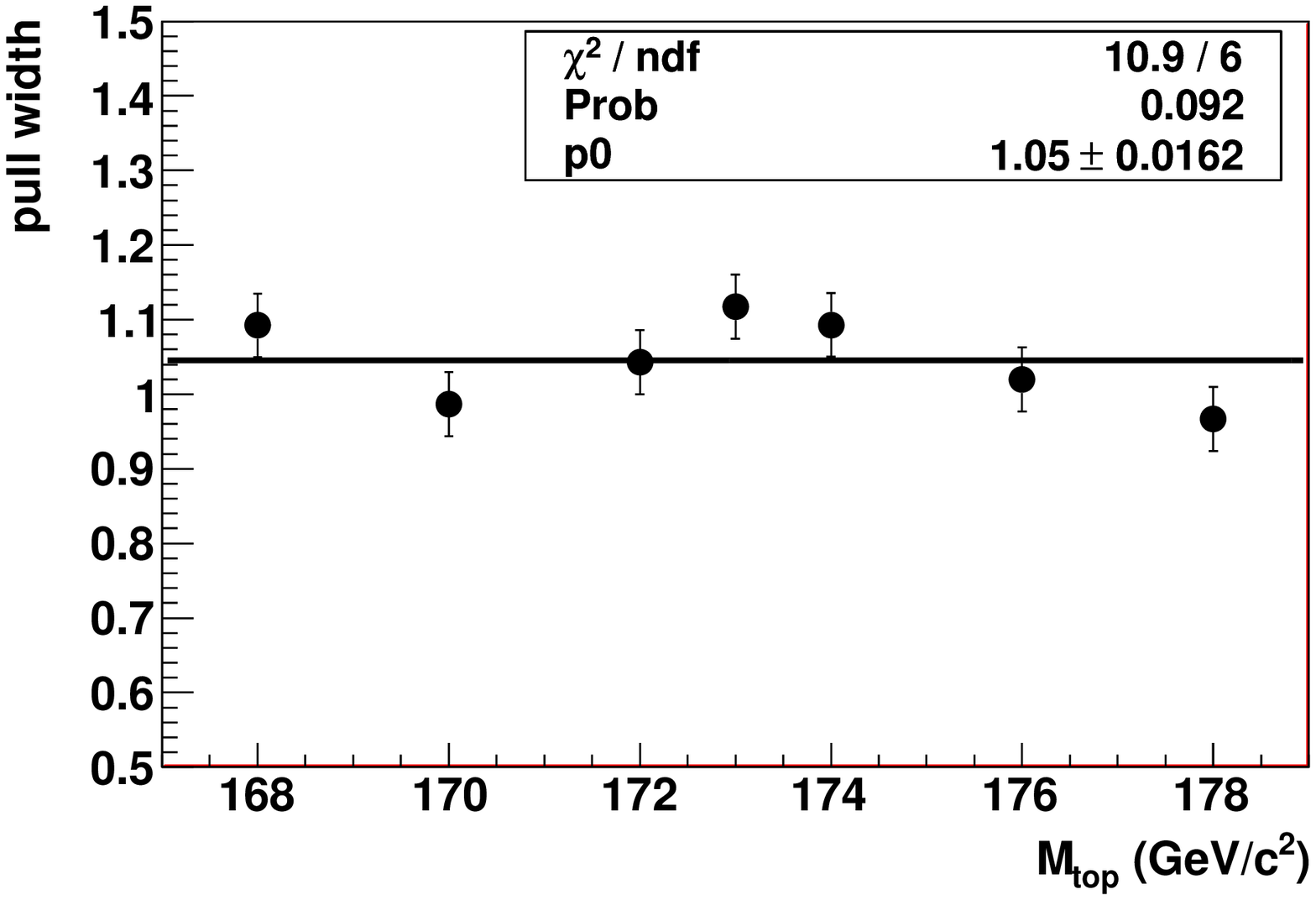} &
\includegraphics[width=0.31\textwidth]{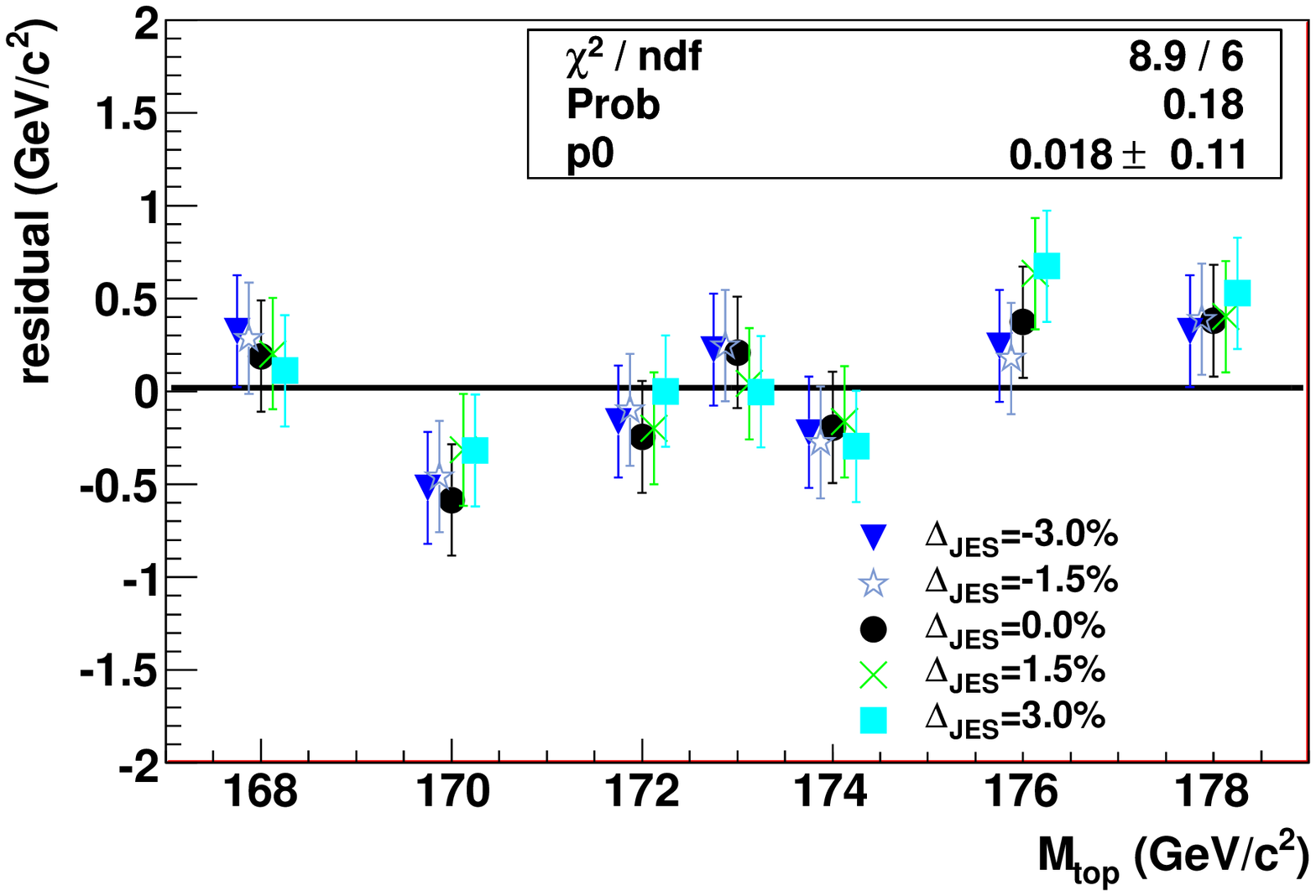} \\
(a) Residual of \mtop & (b) Pull width & (c) Corrected residual\\
\end{tabular}
\caption[variables dependence]{Plots for checking the bias in the fitted top-quark mass (a) and width of the pull distribution (b) for the 2D fit before corrections. With the corrections of the residual and pull width described in the text, we obtain the mass bias for various \djes points at (c) for the 2D fit.}
\label{ref:bias_nocorr}
\end{cfigure1c}

With the corrections of the residual and the width of the pull, we can obtain the expected uncertainty. In Fig.~\ref{ref:err}, the expected statistical uncertainty for the \mtop-only fit~(a) and the 2D fit~(b) have been plotted. Because the 2D fit absorbs the overall uncertainty from JES, it has a larger statistical uncertainty than the \mtop-only fit. If we choose \mtop~=~\gevcc{173}, close to the world average of \mtop~\cite{top_average}, the 2D fit has an expected statistical uncertainty of \gevcc{0.60 \pm 0.03}, whereas that of the \mtop-only fit is \gevcc{0.36 \pm 0.02}. However, the \mtop-only fit may have a much larger JES systematic uncertainty. To obtain the systematic uncertainty of JES, we check the mass residual as a function of \djes, as shown in Fig.~\ref{ref:err} (c)~(\mtop-only fit) and (d)~(2D fit). If we consider an optimistic 2\% overall JES uncertainty~\cite{cmsjes}, the \mtop-only fit gives an expected JES systematic uncertainty of \gevcc{1.68}, whereas that in the case of the 2D fit is \gevcc{0.08}. If we consider a realistic JES systematic uncertainty, which will be 2\%--3\% depending on jet \pt and $\eta$, the \mtop-only fit may have a JES systematic uncertainty slightly larger than \gevcc{1.68}, which is consistent with the recent CMS measurement of \gevcc{2.0} without {\it in situ} JES calibration. However, the JES systematic uncertainty for the 2D fit will still be very small.  We calculate the expected uncertainty from statistics and overall JES systematics together with a quadrature sum. The \mtop-only fit gives an expected uncertainty of \gevcc{1.72}, whereas that in the case of the 2D fit is much smaller, \gevcc{0.61}. The expected statistical uncertainty of \djes in the 2D fit is $0.61 \pm 0.05$\% if we use \mtop~=~\gevcc{173} and a \djes~=~0.0\% sample.

\begin{cfigure1c}
\begin{tabular}{ccc}
\includegraphics[width=0.4\textwidth]{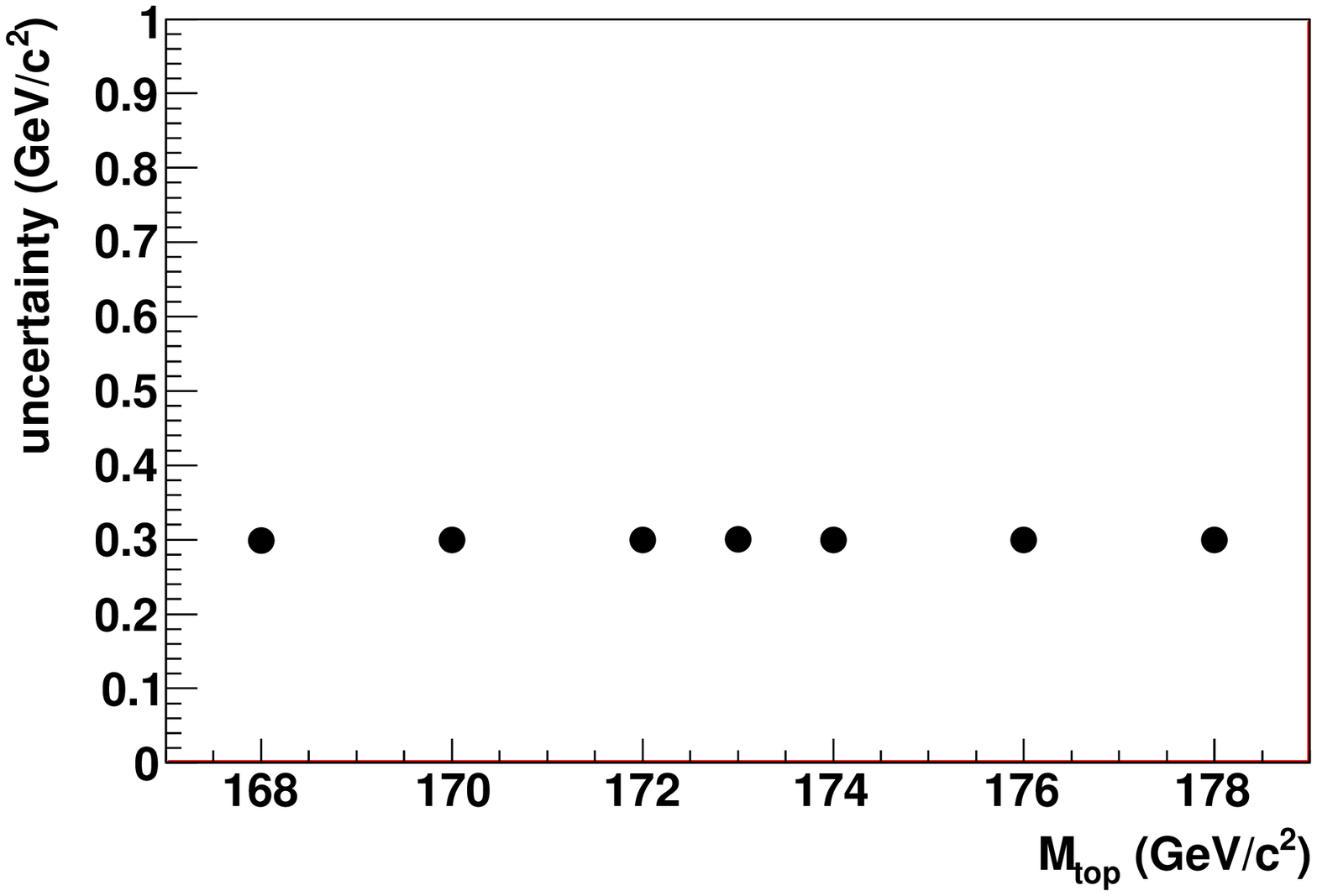} &
\includegraphics[width=0.4\textwidth]{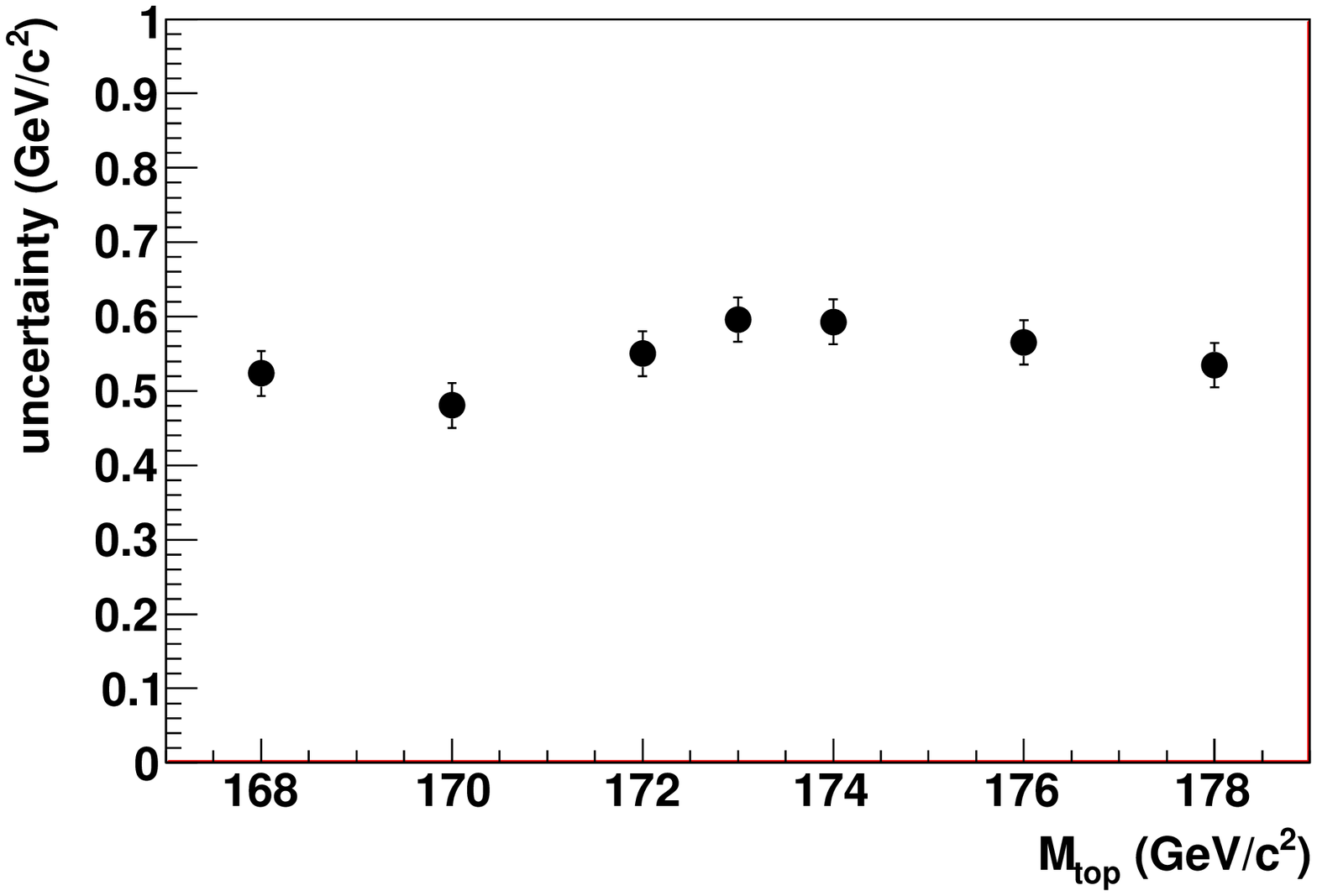} \\
(a) \mtop-only fit  & (b) 2D fit \\
\includegraphics[width=0.4\textwidth]{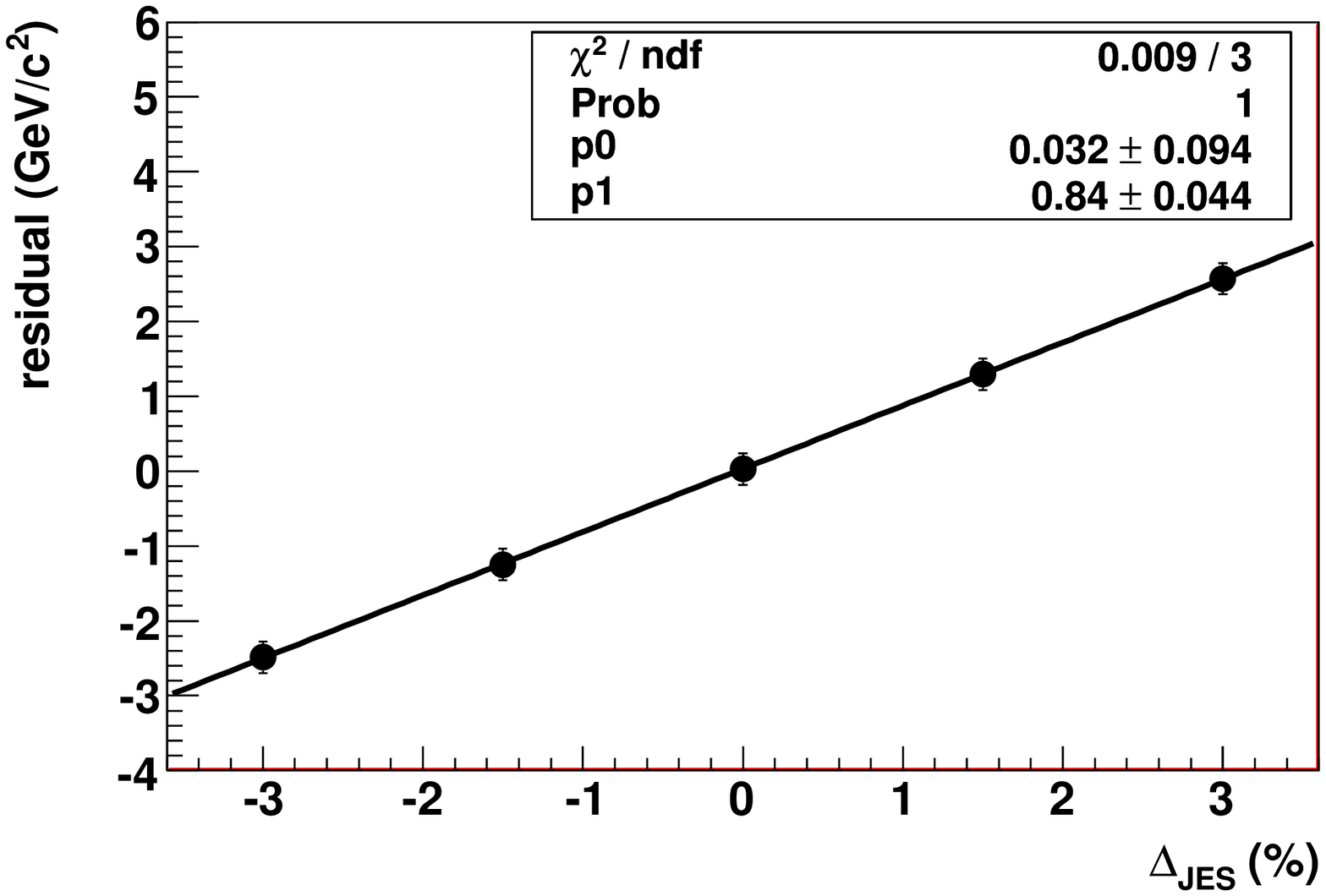} &
\includegraphics[width=0.4\textwidth]{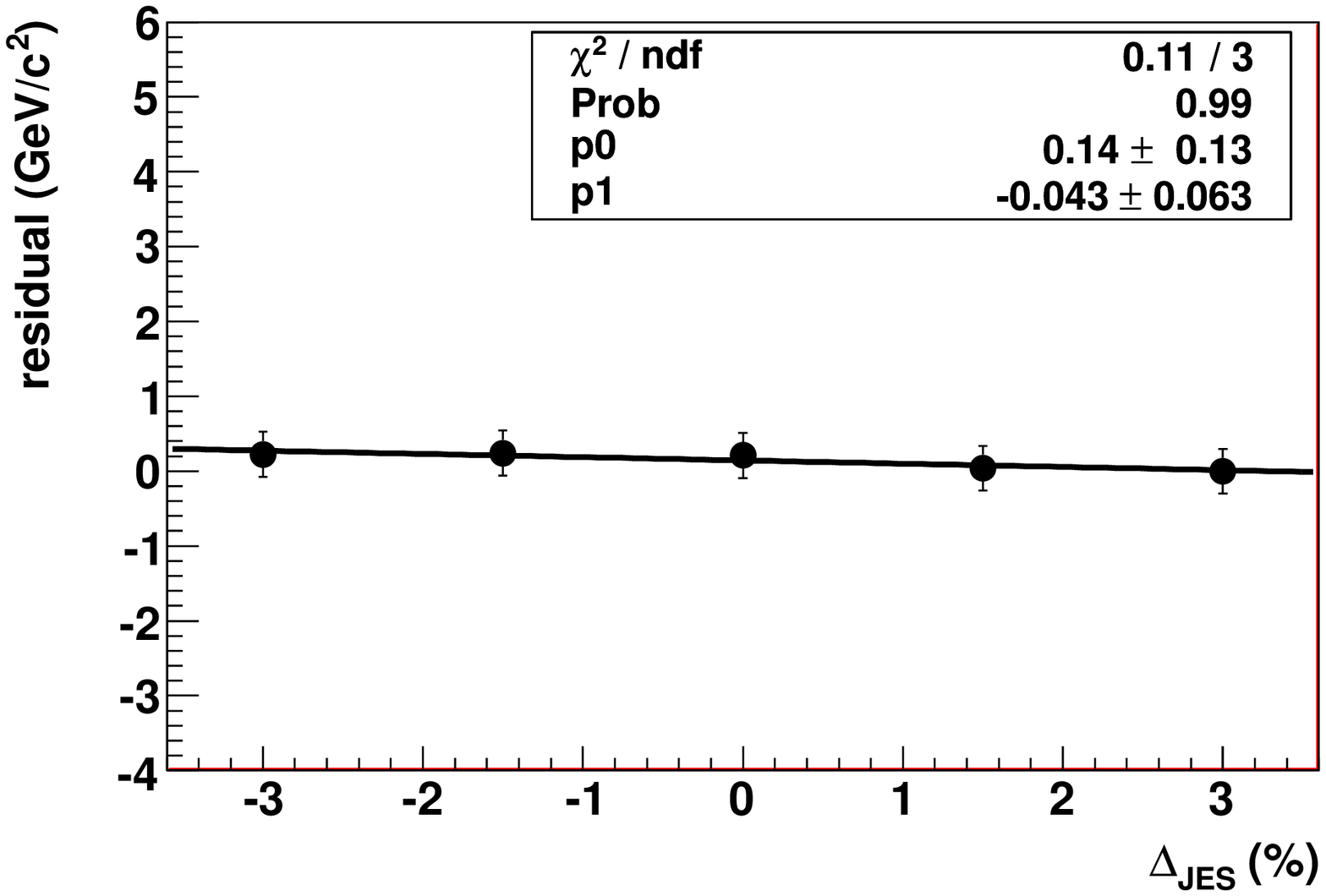} \\
(c) \mtop-only fit& (d) 2D fit \\
\end{tabular}
\caption[uncertainties]{The expected statistical uncertainty for the \mtop-only fit~(a) and the \mtop-\djes 2D fit~(b). To evaluate the systematic uncertainty from the overall JES uncertainty, the mass residuals as a function of \djes are shown for the \mtop=\gevcc{173} sample in (c)~(\mtop-only fit) and (d)~(2D fit).}
\label{ref:err}
\end{cfigure1c}

In conclusion, we have presented a novel technique for  top-quark mass measurement in the dilepton channel by performing {\it in situ} JES calibration using three different variables. Our study shows an approximately \gevcc{0.61} statistical uncertainty including overall JES systematic uncertainty, with \invfb{5} LHC data. This technique significantly improves the precision of the \mtop measurement as compared to the \mtop-only fit. In particular, the overall JES uncertainty, which would be approximately \gevcc{1.7} without additional calibration of JES with a 2\% overall uncertainty assumed, is significantly reduced with the {\it in situ} JES calibration method. To obtain  precision below \gevcc{1}, one still needs to improve other important systematic uncertainties in the LHC experiments. However, the experience at the Tevatron presented a well-controlled systematic uncertainty of around \gevcc{0.86} from the other systematic sources~\cite{massljcdf}. If the other systematic uncertainties are controlled to the level of the Tevatron,  we eventually can reach a precision of \gevcc{1} in the dilepton channel.

This work was supported by a National Research Foundation
of Korea grant funded by the Korean government (NRF-2011-35B-C00007).

\end{document}